\documentstyle[12pt,epsf]{article}

\def\cxx{1 - {{4x^{2} P^{2}}\over{Q^{2}}} }
\def\cx{1 - {{2x    P^{2}}\over{Q^{2}}} }

\def\cut{\sqrt{1 - {{4(m_q^{2}+\lambda^2)}\over {W^2}} } }

\begin{document}

\begin{flushright}
SLAC-PUB-8025\\
December 1998
\end{flushright}

\begin{center}
{\large \bf Gluon Virtuality and Heavy Sea Quark Contributions to the
Spin-Dependent
$g_1$ Structure Function}\\[14 mm]
{\bf Steven D. Bass $^{a,}$}
\footnote{Steven.Bass@mpi-hd.mpg.de; present address: Physik Department,
Technische Universit\"at \break {}\hbox{\hskip .20in} 
M\"unchen, D-85747 Garching, Germany.},
{\bf Stanley J. Brodsky $^{b,}$
\footnote{sjbth@slac.stanford.edu; work supported by the Department of Energy
under contract number \hbox{\hskip .25in}DE--AC03--76SF00515.}
} and {\bf Ivan Schmidt $^{c,}$}
\footnote{ischmidt@fis.utfsm.cl; work supported by Fondecyt (Chile) under
grant 1960536
and by a \break {}\hbox{\hskip .20in} C\'atedra Presidencial (Chile).}
\\[10mm]
{\em $^a$
     Max Planck Institut f\"{u}r Kernphysik, \\
     Postfach 103980,
     D-69029 Heidelberg, Germany}\\[5mm]
{\em $^b$
     Stanford Linear Accelerator Center, \\
     Stanford University, Stanford, California 94309, U.S.A.}\\[5mm]
{\em $^c$
     Departmento de F\'\i sica,
     Universidad T\'ecnica Federico Santa Mar\'\i a, \\
     Casilla 110-V, Valpara\'\i so, Chile}
\end{center}

\vskip 10 mm
\begin{abstract}
\noindent
We analyze the quark mass dependence of photon gluon fusion in polarized
deep inelastic scattering for both the intrinsic and extrinsic gluon
distributions of the nucleon. We calculate the effective number of flavors for
each of the heavy and light quark photon gluon fusion contributions to the
first moment of the spin-dependent structure function $g_1(x)$.
\end{abstract}

\renewcommand{\labelenumi}{(\alph{enumi})}
\renewcommand{\labelenumii}{(\roman{enumii})}

\newpage

\section{Introduction}

One of the most interesting aspects of deep inelastic lepton-proton
scattering is the contribution to the $g_1^p$ spin-dependent structure
function from photon-gluon fusion subprocesses $\gamma^*(q)
g(p)\rightarrow q\bar q$.  Naively, one would expect zero contributions
from light mass $q\bar q$ pairs to the first moment $\int^1_0 dx\,
g_1^p(x,Q^2)$ since the $q$ and $\bar q$ have opposite helicities.  In
fact, this is not the case if the quark mass $m_q$ is small compared to
a scale set by the spacelike gluon virtuality $p^2$.  This is the origin
of the so-called anomalous correction $-3\frac{\alpha_s}{2\pi}\, \Delta
g$ \cite{efremov}-\cite{leader} to the Ellis-Jaffe sum rule \cite{ej} for
isospin-zero targets assuming three light flavors.  Here $\Delta g$ is
the helicity carried by gluons in the hadron target, $\Delta g(Q) =
\int^1_0 dx [g_\uparrow (x,Q) - g_\downarrow(x,Q)]$, at the
factorization scale $Q$.  In the language of the operator product
expansion, the photon-gluon subprocess contributions to the first moment
of $g_1(x,Q)$ correspond to the anomalous VVA triangle graph
\cite{adler,bell} contribution to the hadronic matrix element of the
local axial current.

For fixed gluon virtuality $P^2 = -p^2$ the photon-gluon fusion process
induces two distinct contributions to the first moment of $g_1$ in
polarized deep inelastic scattering.  Let $m_q$ denote the mass of the
struck sea quark.  When $Q^2$ is much greater than both $m_q^2$ and
$P^2$ the box graph contribution to the first moment of $g_1$ for a
gluon target is \cite{bnt}:
\begin{equation}
\int_0^1 dx g_1^{\gamma^{*} g} = - {\alpha_s \over 2 \pi} \left[1 +
\frac{2m_q^{2}}{P^{2}} \frac{1}{\sqrt{1+4m_q^{2}/P^{2}}} \ln \left(
\frac{\sqrt{1+4m_q^{2}/P^{2}} -1}{\sqrt{1+4m_q^{2}/P^{2}} +1}
\right)\right] \ .
\label{eq:1}
\end{equation}
The first, mass-independent term ($- {\alpha_s \over 2 \pi}$) in Eq.
(\ref{eq:1}) comes
from the region of phase space where the struck quark carries large transverse
momentum squared $k_T^2 \sim Q^2$ relative to the photon-gluon direction.
It measures a contact photon-gluon interaction and is associated \cite{ccm}
with the axial anomaly \cite{adler,bell}.  The second mass-dependent term comes from
the region of phase space where the struck quark carries transverse
momentum $k_T^2 \sim P^2, m_q^2$.  This mass dependent term vanishes in
the limit $P^2 \gg m_q^2$ and tends to $+{\alpha_s \over 2 \pi}$ when
$P^2 \ll m_q^2$.  The ``soft'' mass dependent term in Eq.~(\ref{eq:1}) is
associated with the quark parton distribution of the gluon $\Delta
q^{\rm (gluon)}$ ; it can safely be neglected for the light (up and
down) quarks.

On the other hand, the magnitude of the gluon virtuality is important for
gauging the contribution of the massive sea quarks.   If the
sea quark mass is heavy compared to the gluon virtuality $4m^2_q \gg P^2
= - p^2$, the photon-gluon fusion contribution to $\int_0^1 dx\,
g_1(x,Q^2)$ vanishes to leading order in
$\alpha_s(Q^2)$.  This result follows from a general theorem based on
the Drell-Hearn-Gerasimov sum rule \cite{dhg} which states that the
logarithmic integral over the photoabsorption cross section
\begin{equation}
\int ^\infty_{\nu_\pi} \frac{d\nu}{\nu}\
\Delta \sigma_{\gamma a\rightarrow bc}(\nu) = 0(\alpha^3) \ ;
\label{eq:2}
\end{equation}
it vanishes at order $\alpha^2$ for any $2\rightarrow 2$ Standard
Model process \cite{Alt72,Brod95}. Here $\Delta \sigma$ is the
cross section difference for parallel versus anti-parallel incident
helicities.  In the present application, the gluon (for
$p^2 = 0$) takes the role of the on-shell photon $\gamma$ and the particle
$a$ can be taken as a real or virtual photon.  As the photon virtuality
$Q^2$ becomes large, the DHG integral evolves to the first moment of the
helicity-dependent structure function $g_1(x,Q^2)$.   Thus the fusion
$\gamma^* g\rightarrow q\bar q$ Born contribution to $\int_0^1 dx\,
g_1(x,Q^2)$ vanishes for small gluon virtuality $P^2 \ll 4m^2_q$, $P^2
\ll Q^2$.  Notice that the Born photon-gluon fusion contribution to the
Ellis-Jaffe moment is zero even for very light quarks as long as the
gluon virtuality can be neglected.

The above application of the DHG theorem holds for any photon virtuality
$q^2 = -Q^2$, and is thus more general than leading twist
\cite{bassbs}.  In fact, the leading-order fusion contribution to the
$d\nu/\nu$ moment of the difference of helicity-dependent
photo-absorption cross sections vanishes even if $Q^2 < 4m^2_q$,
as long as the gluon virtuality can be neglected.  The result also holds
for the weak as well as electromagnetic current probes
\cite{Brod95,rizzo}.

It is clearly important to ascertain the actual numerical contribution of
heavy quarks $s\bar s$, $c\bar c$, $b\bar b$, $t\bar t$ to the first moment 
of $g_1$; i.e., what is the effective number of sea quark contributions to the
Ellis-Jaffe moment?  From the above discussion, the specific
contribution of a given sea quark pair $q\bar q$ depends not only on
$Q^2$, but more critically on the ratio of scales $p^2/4m^2_q$.
In a full QCD calculation of photon-gluon fusion contributions to the
first moment of $g_1$ one needs to integrate over the distributions of
extrinsic and intrinsic gluon virtualities in the target nucleon.  For
small gluon virtualities ($P^2 \ll m_q^2$) the ``hard'' anomaly
contribution to the first moment of $g_1^{\gamma^* g}$ cancels with the
``soft'' mass dependent contribution.  For deeply-virtual gluons the
mass-independent anomaly contribution dominates over the mass-dependent
term which tends to zero.  Therefore, we shall investigate the effect of
retaining the finite quark masses and performing a more exact analysis,
in which we integrate over $P^2$.  Our aim is to understand the role of
heavy quarks (e.g. strange and charm) in the photon-gluon fusion
process.

As we shall show in the next section, the exact form of the
spectrum $N(p^2)$ of gluon virtuality in the target nucleon
depends in detail on the physics of the nucleon wavefunction.
``Extrinsic'' gluon contributions, which arise from gluon bremmstrahlung
$q_V \rightarrow q_V g$ of a valence quark, have a relatively hard
virtuality $dN_{ext}(p^2)/dp^2 \sim \alpha_s(p^2)/p^2$, above a minimum
virtuality $p_{\rm min}^2$.  The mean virtuality of the extrinsic
gluons depends on the upper limit of integration, which in turn depends on the
kinematic phase space. On the other hand, intrinsic gluons, which are
associated with the physics of the nucleon wavefunction (for example,
gluons emitted by one valence quark and absorbed by another quark), have a
relatively soft spectrum.  We will characterize the shape of the intrinsic
gluon virtuality by the convergent form $dN_{int}(p^2)/dp^2
\sim dN_{ext}(p^2)/dp^2/[1+p^2/{\cal M}^2]$, where ${\cal M}$
is a typical hadronic mass scale.  We shall use such model forms for the
extrinsic and intrinsic gluon distributions to predict specific
contributions of the heavy sea quarks to the first moment of
$g_1(x,Q^2)$.

In addition to the photon-gluon fusion contributions, additional
contributions to the first moment of $g_1(x,Q^2)$ arise from intrinsic
heavy sea quarks associated with higher Fock states in the target
hadron.  For example, meson-baryon fluctuations such as $p \rightarrow K
\Lambda$ imply a negative intrinsic strange quark contribution to
$\int_0^1 dx\, g_1(x,Q^2)$ \cite{negative}.  In the case of charm, the
small probability 0(1\%) of intrinsic charm present in the proton implies
a small intrinsic charm contribution to $\int_0^1 dx\, g_1(x,Q^2)$.

The charm contribution to the nucleon helicity-dependent structure
functions and sum rules will be addressed by several new experiments.
The COMPASS \cite{compass} and HERMES \cite{hermesc} experiments 
will measure charm production \cite{guillet}-\cite{neervan}
in polarized deep inelastic scattering. Experiments have also been
proposed at SLAC \cite{bosted}.  The aim of these experiments is to learn 
about the gluon polarization in a nucleon through the photon-gluon fusion process.

\section{Polarized gluons and $g_1$}

In order to analyze the sensitivity of the anomaly to the sea quark mass
in the photon-gluon fusion subprocesses, let us start by expressing the
contributing gluon distributions in terms of the corresponding
bound-state wavefunctions.  In general the $Q^2$ dependence of the
parton distributions comes from the integral of the bound-state
wavefunction over the virtuality of the corresponding parton up to the
scale $Q^2$.  Schematically, for the polarized gluon distribution, we
have
\begin{equation}
\triangle G(x,Q^{2})\ =\ \int \nolimits
^{Q^{2}}dP^{2}\big[\big\arrowvert
\Psi_{g\uparrow/p\uparrow}(P^{2},x)\big\arrowvert ^{2}-\big\arrowvert
\Psi_{g\downarrow/p\uparrow}(P^{2},x)\big\arrowvert ^{2}\big],
\label{eq:3}
\end{equation}
which means that
\begin{equation}
{\partial \over
\partial P^{2}}\Delta G(x,P^{2})\ =\ \big\arrowvert
\Psi_{g\uparrow/p\uparrow}(P^{2},x)\big\arrowvert ^{2}-\big\arrowvert
\Psi_{g\downarrow/p\uparrow}(P^{2},x)\big\arrowvert ^{2} \equiv
{d^2\Delta N_{g/p}\over dP^{2}dx}.
\label{eq:4}
\end{equation}
Here $\Psi_{g\uparrow/p\uparrow}$ and $\Psi_{g\downarrow/p\uparrow}$ are the
gluon wavefunctions for positive and negative helicities relative to the
proton helicity as functions of the gluon virtuality $P^2= -p^2$ and
the fraction $x$ of the plus component of the target nucleon's momentum.

In perturbative QCD the total photon-gluon fusion contribution to $g_1$
for a nucleon target is given by
\begin{equation}
g_1^{(G)} (x, Q^{2})\
= {1 \over 2} \sum \limits _{q} e_q^2 \ g_1^{(Gq)}(x, Q^2)\ ,
\label{eq:5}
\end{equation}
where $g_1^{(Gq)}$ is the contribution where the struck
quark carries flavor $q$
\begin{equation}
g_1^{(Gq)} (x,Q^{2})\ =\ \int
_{P_{\min}^{2}}^{Q^{2}}dP^{2}\ {\partial (\Delta G(x,P^{2}))\over
\partial P^{2}}\otimes {\cal A}_q (x,Q^{2},P^2) \ .
\label{eq:6}
\end{equation}
Here $\otimes$ denotes the convolution over $x$ and ${\cal A}_q$ denotes the
contribution to the spin structure function $g_1$ of a ``gluon target''
with virtuality $P^2$, where the struck quark carries flavor $q$.  The
infra-red cut-off $P_{\min}^2$ is the minimum gluon virtuality at which
we can apply perturbative QCD---that is, where the current-quark and
gluon degrees of freedom in perturbative QCD give way to dynamical
chiral symmetry breaking and confinement.  The GRV \cite{grvb} and Bag
model \cite{bag} analyses of deep inelastic structure functions involve
taking a QCD-inspired model-input for the leading twist parton
distributions at some low scale $\mu_0^2$, evolving the distributions to
deep inelastic $Q^2$ and comparing with data.  The optimal GRV and Bag
model fits to deep inelastic data are found with $\mu_0^2 \simeq$ 0.2 -
0.3 GeV$^2$.  Motivated by this phenomenological observation, we shall
set $P_{\min}^2 =0.3$GeV$^2$.

In the Born approximation ${\cal A}_q$ is calculated from the box graph
contribution to photon-gluon fusion.
We can define the ``hard'' part of ${\cal A}_q$ by imposing a cutoff on
the transverse momentum squared of the struck quark
$k_T^2 > \lambda^2$ \cite{bnt}:
\begin{eqnarray}
&&{\cal A}_q |_{\rm hard} \big(x,Q^{2},P^{2},\lambda^2\big) =
-{\alpha_s  \over 2 \pi }
{\cut \over \cxx} \Biggl[ (2x-1)(\cx) \\ \nonumber
& &
\biggl(1 - {1 \over {\cut \sqrt{\cxx} }}
\ln \biggl({ {1+\sqrt{\cxx} \cut}\over {1-\sqrt{\cxx} \cut}}
\biggr) \biggr) \\ \nonumber
& &
+ (x-1+{{x P^{2}}\over{Q^{2}}})
{{\left( 2m_q^{2}(\cxx)- P^{2}x(2x-1)(\cx)\right)}
\over {(m_q^{2} + \lambda^2) (\cxx) - P^{2}x(x-1+{{x P^{2}}
\over{Q^{2}}})}} \Biggr] \ .
\label{eq:7}
\end{eqnarray}
Here $m_q$ is the fermion mass, $x$ is the Bjorken variable
and $W^2 = Q^2 ( {1 - x \over x} ) - P^2$
is the center of mass energy for the photon-gluon collision.
The running coupling, $\alpha_s$, in Eq. (7) is evaluated at
the scale $P^2$.
Following Parisi and Petronzio \cite{parisi} we shall use a modified
running $\alpha_s(P^2)$---see Eq. (15) below---which freezes
in the infrared, to describe ${\cal A}_q$ when $P^2$ becomes small.

Keeping contributions where the struck quark carries transverse momentum
squared $k_T^2 \geq \lambda^2$, the photon-gluon fusion contribution to
the first moment of $g_1^{(Gq)}$ is obtained from (6):
\begin{equation}
\Gamma_q (Q^2,\lambda^2)= \ \int_{P_{\min}^{2}}^{Q^{2}}dP^{2}\ \ {\cal
I}_q (P^{2},\lambda^2)
\ {d \Delta N_{g/p}\over dP^{2}}(P^2) \  .
\label{eq:8}
\end{equation}
Here \begin{equation} {\cal I}_q
(P^2,\lambda^2) = \int_0^{x_{\max}} dx \ {\cal A}_q (x, Q^2,
P^2,\lambda^2)
\label{eq:9}
\end{equation}
and
\begin{equation}
{d \Delta
N_{g/p}\over dP^{2}}(P^2) \ =\ \int _{0}^{z_{\max}(P^2)} dz\ {d^2 \Delta
N_{g/p}\over dP^{2}dz}(z,P^2) \ .
\label{eq:10}
\end{equation}
The cutoffs $x_{\max}$ and
$z_{\max}$ in Eqs. (\ref{eq:9},{\ref{eq:10}) come from the kinematics.  For
the box graph
term ${\cal A}_q$, the cutoff $x_{\max} =
Q^2/(Q^2+P^2+4(m_q^2+\lambda^2))$ in Eq.  (\ref{eq:9}) is obtained from the
phase
space factor $\sqrt{1 - {4 (m_q^2 + \lambda^2) \over W^2}}$ in Eq.
(\ref{eq:7}).
The cutoff $z_{\max}$ in Eq.  (\ref{eq:10}) is derived from the explicit form of
the polarized gluon distribution---see below.

In the rest of this Section we discuss the contribution of $q \bar q$
pairs with small transverse momentum (when we relax the $\lambda^2$
cut-off), the size of higher-twist contributions to the first moment of
$g_1^{(Gq)}$, and the jet signature of the different contributions to
the first moment.

When we integrate over the full range of possible impact parameters we
need to include small values of $k_T^2$ in Eqs.  (\ref{eq:8}--\ref{eq:10}).
This necessarily involves extrapolating the calculation into the domain of
non-perturbative QCD.
Shore and Veneziano \cite{shore} have considered the analogous process
of the spin structure function of the polarized photon for a virtual
photon target.  They argue that the target photon virtuality where
$\int_0^1 dx g_1^{\gamma}$ grows from zero (at $P^2=0$) to $- {\alpha
\over \pi} N_c$ depends on the realization of chiral symmetry breaking
in QCD.  In perturbative QCD the individual quark flavor contributions
to $\int_0^1 dx g_1^{\gamma}(x,Q^2)$ grow rapidly from zero when $P^2
\sim 4m_q^2$.  In full QCD, spontaneous chiral symmetry breaking means
that the scale of the transition virtuality is set by the constituent
quark mass rather than by the current quark mass --- that is, we expect
$\int_0^1 dx g_1^{\gamma}(x,Q^2)$ to grow rapidly from zero when $P^2
\sim m_{\rho}^2$.  Motivated by this result, one might expect the
gluon-virtuality where ${\cal I}_q$ grows rapidly to depend on any
possible diquark structure of the gluon at low $k_T^2$.  Spontaneous
chiral symmetry breaking is a considerably more dramatic effect in the
light quark masses than it is in the heavy quark masses. Thus we can
expect that perturbative QCD will provide a reasonable
model-independent estimate of the heavy-quark ${\cal A}_q$ when
$\lambda^2$ becomes small.

When $Q^2 \rightarrow \infty$ the expression for ${\cal A}_q$ simplifies
to the leading twist (=2) contribution:
\begin{eqnarray} 
{\cal A}_q (x,Q^2,P^2,\lambda^2) 
&=& {\alpha_s \over 2 \pi} \Biggl[ (2x-1) \Biggl(
\ln {Q^2 \over \lambda^2} + \ln {1-x \over x} - 1 \Biggr) \\[1ex] 
\nonumber &\qquad& + (2x-1) \ln {\lambda^2 \over {x(1-x) P^2 
+ (m_q^2 + \lambda^2)} }\\[1ex]
&\qquad&+ (1 -x) { {2m_q^2 - P^2x(2x-1)} \over {x(1-x)P^2 
+ m_q^2 + \lambda^2} } \Biggr] \nonumber
\label{eq:11}
\end{eqnarray}
which has the first moment
\begin{equation}
{\cal I}_q (P^2,\lambda^2)
\ = \ - {\alpha_s \over 2 \pi} \left[1 + \frac{2m_q^{2}}{P^{2}}
\frac{1}{\sqrt{1+4(m_q^{2}+\lambda^2)/P^{2}}} \ln \left(
\frac{\sqrt{1+4(m_q^{2}+\lambda^2)/P^{2}} -1}
{\sqrt{1+4(m_q^{2}+\lambda^2)/P^{2}} +1} \right)\right].
\label{eq:12}
 \end{equation}
For finite quark masses, the cutoff $x_{\max}$ protects ${\cal A}_q$ from
reaching the $\ln (1-x)$ singularity in Eq. (\ref{eq:7}).  To quantify this
effect, in Table \ref{tab:1} we list the values of ${\cal I}_q$ for different
values of $P^2$ and $\lambda^2$.  The ``$Q^2 = \infty$'' values are
obtained by keeping only the leading twist contribution, Eq. (\ref{eq:13}).  For
the ``hard'' cut-off $\lambda^2 =1 $\ GeV$^2$ the cut-off itself acts as a
major source of higher-twist.  When we relax the cut-off by setting
$\lambda^2$ to zero we find a large $\simeq 63\%$ higher twist
suppression of ${\cal I}_c$ at $Q^2=10 $\ GeV$^2$.  Steffens and Thomas
\cite{steffens} have observed that ${\cal I}_c$ is, to good
approximation, independent of $P^2$ for values of $P^2$ between zero and
1 GeV$^2$.  The rise in ${\cal I}_c$ with increasing $\lambda^2$
corresponds to removing a greater amount of the mass dependent term in
Eq.  (\ref{eq:1}) which cancels against the mass-independent (anomaly) term
which is associated with $k_T^2 \sim Q^2$ in the limit $Q^2 \rightarrow
\infty$.  When we increase the cut-off $\lambda^2$ on the transverse
momentum squared of the struck quark we increase the infrared cut-off on
the invariant mass of the $q {\overline q}$ pairs produced by the
photon-gluon fusion.  Quark mass dependent contributions to ${\cal I}_q$
go to zero when we increase the invariant mass ${\cal M}_{q {\overline
q}}$ much greater than $4 m_q^2$.  Note that the light-quark ${\cal
I}_l$ in Table \ref{tab:1} is significantly suppressed below unity with the
$\lambda^2 = 1 $\ GeV$^2$ cut-off at $Q^2=10 $\ GeV$^2$.  If we decrease the
cut-off this ${\cal I}_l$ grows to 0.87 ($\lambda^2 =0.5 $\ GeV$^2$), 0.91
($\lambda^2 =0.3 $\ GeV$^2$) and 0.96 ($\lambda^2 =0.1 $\ GeV$^2$) when we
use our modified $\alpha_s$ (---see Eq. (\ref{eq:15}) below) together with the
current light-quark mass through-out.

\begin{table}
\begin{center}
\caption{Heavy quark effects in ${\cal I}_q$
(in units of $-{\alpha_s (P^2) \over 2 \pi}$) }
\label{tab:1}
\begin{tabular} {cccccc}
\\
\hline\hline
\\
$P^2$  &  $\lambda^2$  &  $Q^2$  &  light  &  strange  &  charm  \\
\\
\hline
 0.5   &  1.0  &    10.0  &  0.77  &  0.74  &  0.16  \\
 0.5   &  1.0  &   100.0  &  0.98  &  0.94  &  0.30   \\
 0.5   &  1.0  & $\infty$ &  1.00  &  0.96  &  0.32    \\
\hline
 0.5   &  0.0  &   10.0   &  0.99  &  0.61  &  0.013    \\
 0.5   &  0.0  &  100.0   &  1.00  &  0.63  &  0.033     \\
 0.5   &  0.0  & $\infty$ &  1.00  &  0.63  &  0.035      \\
\hline\hline
\end{tabular}
\end{center}
\end{table}

Consider the large $Q^2$ limit ($Q^2 \gg 4 m_q^2$).  When we set
$\lambda^2=0$ in Eq. (\ref{eq:12}) to obtain Eq. (\ref{eq:1}), the ``hard''
anomaly contribution to ${\cal I}_q$ (the first moment of $g_1^{\gamma^* g}$)
cancels with the ``soft'' mass dependent contribution for small gluon
virtualities ($P^2 \ll m_q^2$).  For deeply virtual gluons the
mass-independent anomaly contribution dominates over the mass-dependent
term which tends to zero in the limit $P^2 \gg 4 m_q^2$.  It is
interesting to observe that in a semi-inclusive experiment one can in
principle identify events which correspond specifically to the
contributions to the first moment of $g_1(x)$.  These are events with
three jets recoiling, taking up the large momentum $q_T$ transferred by
the lepton.  The final state consists of the $q\bar q$ pair with
perpendicular momentum $\vec q_T- \vec p_T$ plus the quark that emitted
the gluon, with mass $M$ and transverse momentum $\vec p_T$.  These
events have gluon virtuality $-p^2 \geq m^2_q$.  As shown in
\cite{brod97}, the events where only the $q \bar q$ pair recoils produce
no contribution to the first moment of $g_1(x)$.  This corresponds to
events with small transverse momentum of the quark that emitted the
gluon $p_T^2 \ll m_q^2$, or with gluon virtuality $-p^2 \ll m^2_q$.  The
vanishing of the first moment of $g_1(x)$ for heavy quark production
implies that there must be a polarization asymmetry zero, which in
principle can be measured experimentally \cite{brod97}.

\section{Extrinsic and intrinsic glue}

Having established the theoretical framework, we now investigate
photon-gluon fusion using simple models for the exclusive and
inclusive gluon distributions of the nucleon.

The ``extrinsic'' glue consists of gluons which are radiatively
generated from individual valence quarks in the target whereas the
``intrinsic'' glue is associated with gluon exchange between valence
quarks.  For example, consider a gluon which is exchanged between two
valence quarks in the proton.  In constituent quark models these ``gluon
exchange currents'' contribute to the proton-Delta mass difference
\cite{myhrer}.  They also renormalize the valence contributions to the
nucleon's axial charges, which are measured in $\beta-$decays and in the
first moment of $g_1(x,Q^2)$.  When cut, the exchanged gluon gives an
intrinsic gluon.  A gluon which contributes to the quark self energy
when cut gives an extrinsic gluon.

We now estimate the size of the extrinsic and intrinsic gluon
contributions to the first moment of $g_1$.  We calculate the ratio
$n_{\rm eff} = \Gamma_h / \Gamma_l$ of the heavy to light quark
contributions to photon gluon fusion for both the extrinsic and
intrinsic glue.

The extrinsic and intrinsic gluon distributions are dominated by gluons
with small virtuality.  The virtuality distribution ${d N_{g/p} \over
dP^2}$ for the extrinsic glue contains a logarithmic tale extending to
the kinematic limits.  Phenomenologically \cite{brodin}, the momentum
distribution of the intrinsic glue is found to be weighted by the factor
\begin{equation}
{\cal W}(P^2) = N_g \biggl( {1 \over 1 + {P^2 \over
{\cal M}^2}} \biggr)
\label{eq:13}
\end{equation}
relative to the extrinsic glue.  The mass parameter ${\cal M}^2$ can be
estimated as 0.71 GeV$^2$, the mass scale of the dipole fit to the
proton form factor \cite{brodin}; $N_g$ is a model dependent
normalization constant.

We start with a simple model for the gluon distributions taking into
account their correct $P^2$ distribution.  Model dependent normalization
uncertainties cancel in the ratios $\Gamma_h / \Gamma_l$.  We treat the
nucleon target as a three-quark system where the target quarks are
treated as ``elementary'' with constituent quark mass $M$ equal to 300
MeV. The polarized extrinsic gluon distribution is given by
\begin{equation}
{d^2\Delta N_{g/p}\over dP^{2}dz}(z,P^2)\ =\ {\cal N} C_F\
{\alpha_s \over 2\pi}
\left(1\over
z\right){{(1-(1-z)^{2})p_{T}^{2} + M^2 z^4}\over
(p_{T}^{2}+M^{2} z^{2})^2}
\ (1-z)
\label{eq:14}
\end{equation}
where the $(1-z)$ factor is a Jacobian factor for the change of
variables from $p_T^2$ to $P^2$ in Eq. (\ref{eq:8}).  The QCD 
factor $C_F = {4 \over 3}$; ${\cal N} \simeq 0.6$ 
\cite{cloudy,weise,schlumpf} is the spin depolarization factor found
in relativistic quark models, which parametrizes the transfer of the 
proton's angular momentum from intrinsic spin of the quarks to 
orbital angular momentum through relativistic effects and quark-pion 
coupling.  Since the gluon transverse momentum squared 
$p_{T}^{2}=P^{2}(1-z)-M^{2}z^{2}$ is
non-negative, we obtain the $z_{\max}$ cutoff in Eq. (\ref{eq:10}):
$z_{\max}(P^{2})=(-1+\sqrt {1+4C})/2C$, where $C=M^2/P^2$.

Our simple model, Eq. (\ref{eq:14}), for the gluon distributions exhibits the $x
\rightarrow 0$ behavior predicted by color coherence \cite{gluon}.  By
construction, it also exhibits the large $x$ behavior associated with an
elementary quark target.  In a more sophisticated model one should also
include gluon exchange between the valence quarks in addition to the
gluon involved in the $\gamma^* g \rightarrow q {\overline q}$ process.
However, this is beyond the scope of the present paper.

We work in the analytically extended \cite{extms} $\alpha_V$ scheme
\cite{alphav}.  This means that we use the running coupling
\begin{equation}
\alpha _{s}(P^{2})\ =\ {4\pi \over \beta _{0}\ln
({P^{2}+4m_{g}^{2}\over \Lambda _{V}^{2}})},
\label{eq:15}
\end{equation}
in Eqs. (7,14).
Here $m_g^2 = 0.2 $GeV$^2$, $\Lambda_V = 0.16 $GeV and the
number of flavors which contribute to $\beta_0$ is taken as a continuous
variable which depends on $P^2$ \cite{extms}:  $\beta _{0}=(11-{2\over
3}\sum \limits _{i=1}^{4}N_{i})$ where
\begin{equation}
N_i \simeq \biggl(1 +
{5 \over \rho_i} \biggr)^{-1}, \ \ \ \ \ \ \ \ (\rho
_{i}=P^{2}/m_{i}^{2}) \ .
\label{eq:16}
\end{equation}

\begin{figure}[htbp] 
\begin{center}
\leavevmode 
{\epsfbox{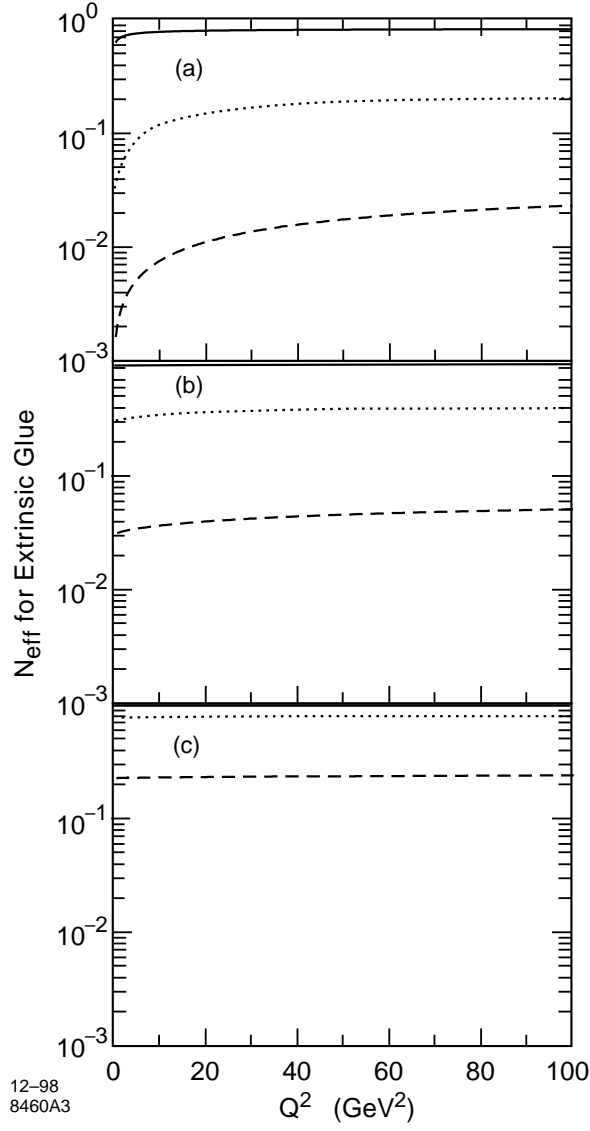}}
\end{center}
\caption[*]{The effective number of flavors $n_{\rm eff}$ for heavy sea 
quarks $s \bar s, c \bar c, $ and $b \bar b$ contributing to the first moment  
of $g_1(x,Q^2)$, arising from $\gamma^*$-(extrinsic gluon) fusion,  
as a function of  momentum transfers  $Q^2 < 100 $GeV$^2$. In Fig. 
\ref{fig1}a,  the cutoff  on quark transverse momentum 
$k_T^2 > \lambda^2$ is set equal to zero. In Figs. \ref{fig1}b 
and \ref{fig1}c,  $\lambda^2=$1  GeV$^2$  and $\lambda^2 =$ 10 
GeV$^2$, respectively.}
\label{fig1}
\end{figure}

In Figs. \ref{fig1}a--\ref{fig1}c we show the effective number of flavors, 
$n_{\rm eff} = \Gamma_h / \Gamma_l$, for the heavy flavor $s \bar s$, 
$c \bar c$ and $b \bar b$ production contributions to the first moment 
of $g_1(x,Q^2)$ from $\gamma^*$-(extrinsic gluon) fusion up to 
$Q^2=100 $GeV$^2$.  Figure \ref{fig1}a is obtained by setting 
$\lambda^2$ equal to zero.  (In this calculation we have used 
``modified'' $\alpha_s$, Eq.  (\ref{eq:15}), together
with the current light-quark mass through-out.)  In Figs. \ref{fig1}b 
and \ref{fig1}c we set the cut-off $\lambda^2$ equal to 1 GeV$^2$ 
and 10 GeV$^2$
respectively.  We repeat these calculations for $\gamma^*$-(intrinsic
gluon) fusion in Figs.\ref{fig2}a--\ref{fig2}c.  The results in Figs. \ref{fig1} 
and \ref{fig2} are summarized in Table \ref{tab:2}.
\begin{table}
\begin{center}
\caption{The effective number of heavy-flavors
$n_{\rm eff} = \Gamma_h / \Gamma_l$. }
\label{tab:2}
\begin{tabular} {ccccc}
\\
\hline\hline
\\
$Q^2$ & $\lambda^2$ & $n_{\rm s}$ & $n_{\rm c}$ & $n_{\rm b}$ \\
\\
\hline
\\
Extrinsic glue & & & & \\
\hline
 10.0  &   0.0  &  0.78  &  0.12  &  0.007  \\
 10.0  &   1.0  &  0.97  &  0.35  &  0.037  \\
 10.0  &  10.0  &  1.00  &  0.78  &  0.23   \\
\hline
100.0  &   0.0  &  0.82  &  0.21  &  0.023  \\
100.0  &   1.0  &  0.97  &  0.41  &  0.052  \\
100.0  &  10.0  &  1.00  &  0.79  &  0.24   \\
\hline
Intrinsic glue & & & & \\
\hline
 10.0  &   0.0  &  0.71  &  0.07  &  0.004  \\
 10.0  &   1.0  &  0.97  &  0.33  &  0.034  \\
 10.0  &  10.0  &  1.00  &  0.78  &  0.23   \\
\hline
100.0  &   0.0  &  0.71  &  0.08  &  0.005  \\
100.0  &   1.0  &  0.97  &  0.33  &  0.035  \\
100.0  &  10.0  &  1.00  &  0.78  &  0.23   \\
\hline\hline
\end{tabular}
\end{center}
\end{table}
\begin{figure}[hbtp] 
\begin{center}
\leavevmode
{\epsfbox{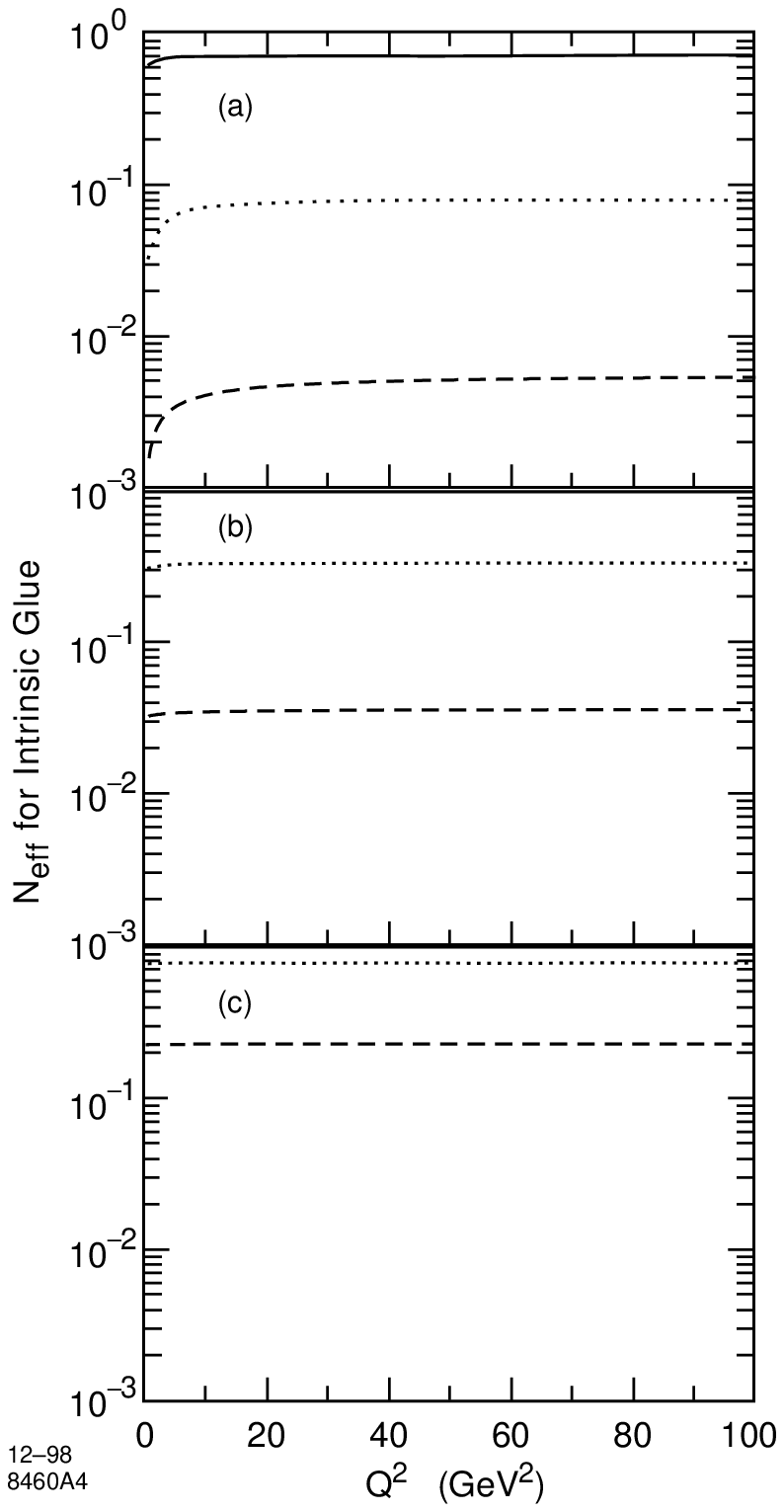}}
\end{center}
\caption[*]{The effective number of flavors $n_{\rm eff}$ for heavy sea 
quarks $s \bar s, c \bar c, $ and $b \bar b$ contributing to the first moment  
of $g_1(x,Q^2)$, arising from $\gamma^*$-(intrinsic gluon) fusion. In 
Fig. \ref{fig2}a,  the cutoff  on quark  tranverse momentum $k_T^2 > 
\lambda^2$ is set equal to zero. In Figs. \ref{fig2}b and \ref{fig2}c,  
$\lambda^2=$1 GeV$^2$ and $\lambda^2  =$ 10 GeV$^2$, respectively. }
\label{fig2}
\end{figure}

\begin{figure}[hbt] 
\begin{center}
\leavevmode
{\epsfbox{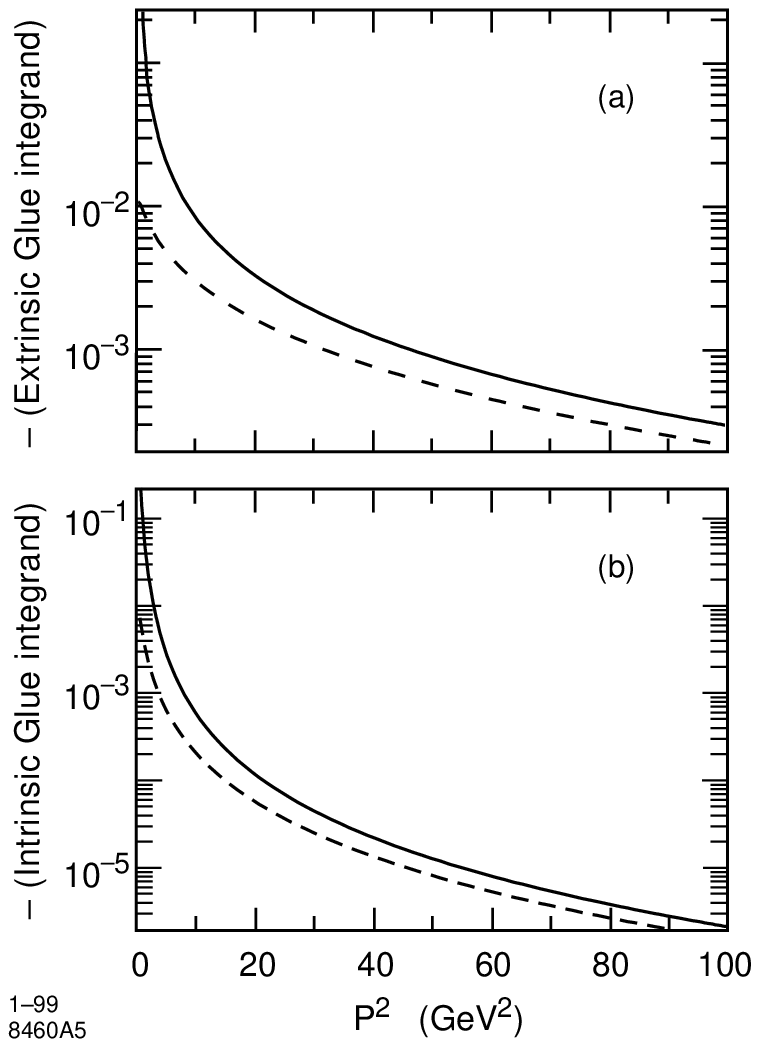}}
\end{center}
\caption[*]{The integrand in Eq. (\ref{eq:8}) as a function of $P^2$ for
the light and charm quark contributions at $Q^2=100~$GeV$^2$ for
extrinsic glue.  Figure \ref{fig3}b shows the corresponding integrand for
intrinsic glue.}
\label{fig3}
\end{figure}

The effective number of flavors $n_{\rm eff} = \Gamma_h / \Gamma_l$
increases for the heavy quarks when we increase the cutoff $\lambda^2$
on the transverse momentum squared of the struck quark.  This
corresponds to removing a greater amount of the mass dependent term in
Eq.  (\ref{eq:1}) which cancels against the mass-independent (anomaly) term from
$k_T^2 \sim Q^2$ in the limit $Q^2 \rightarrow \infty$.  By increasing
$\lambda^2$ we are increasing the cut on the invariant mass of the $q
{\overline q}$ pairs produced by the photon-gluon fusion.  Quark-mass
dependent terms become less important when the invariant mass ${\cal
M}_{q {\overline q}}$ becomes much greater than $4 m_q^2$.

In Fig. \ref{fig3}a we show the integrand in Eq. (\ref{eq:8}) as a function 
of $P^2$ for the light and charm quark contributions at $Q^2=100~$GeV$^2$ 
with the extrinsic glue.  Figure \ref{fig3}b shows the corresponding integrand 
for the intrinsic glue.  Figures \ref{fig3}a and \ref{fig3}b both involve 
$\lambda^2=0$.  Note that the heavy and light quark curves come closer 
together with increasing $P^2$.  This result corresponds to the fact that the 
mass-dependent term in Eq. (\ref{eq:1}) tends to zero in the limit $P^2 \gg 4 m_q^2$.

\section{Phenomenology and discussion}

Gluon polarization offers a possible explanation for the small value of
$g_A^{(0)}$ (the three-flavor, singlet axial charge) extracted from
polarized deep inelastic scattering \cite{expt,smc,hermes,slac}:
\begin{equation}
g_A^{(0)} \simeq 0.2 - 0.35 .
\label{eq:17}
\end{equation}
Relativistic binding \cite{schlumpf} and constituent-quark pion coupling
\cite{cloudy,weise} models predict $g_A^{(0)} \simeq 0.6$ --- a factor
of two larger than the measured $g_A^{(0)}$.  In these semi-classical
models $g_A^{(0)}$ is interpreted as the fraction of the nucleon's
helicity which is carried by its quark constituents.  In QCD the axial
anomaly \cite{adler,bell} induces various gluonic contributions to
$g_A^{(0)}$.  One finds \cite{efremov,ar,ccm,bass98}
\begin{equation}
 g_A^{(0)} = \Biggl( \sum_q \Delta q - 3 {\alpha_s \over 2
\pi}
\Delta g
\Biggr)_{\rm partons} + \ {\cal C}  \ .
\label{eq:18}
\end{equation}
Here ${1 \over 2}
\Delta q$ and $\Delta g$ are the amount of spin carried by quark and
gluon partons in the polarized proton. The $- 3 {\alpha_s \over 2 \pi}
\Delta g_{\rm partons}$ term is associated with the mass-independent,
local $\gamma^{*} g$ interaction in Eq. (\ref{eq:1}) assuming three light
flavors. The soft mass-dependent contributions to photon-gluon fusion are
included in $\Delta q_{\rm partons}$.  The last term, ${\cal C}$, is associated with
non-trivial  gluon topology \cite{bass98} and a possible $\delta (x)$
term in $g_1$.   It is missed by polarized deep inelastic scattering
experiments which measure the combination $(g_A^{(0)} - {\cal C})$.

How large  are the photon-gluon fusion sea-quark contributions $\Gamma_q$
to $g_A^{(0)}$ if we allow for finite sea-quark masses and a spectrum of
gluon virtuality?

Since extrinsic glue is radiatively generated from single quark lines in
the target, we believe that our model should provide a
good order-of-magnitude estimate for the normalization of the extrinsic
$\Gamma_q$. We find $\Gamma_c^{\rm ext} = -0.0024$ and $(\Gamma_u +
\Gamma_d + \Gamma_s)^{\rm ext} = -0.033$ at $Q^2 = 100~$GeV$^2$ 
for the extrinsic glue.

The magnitude of the intrinsic gluon contribution to $g_A^{(0)}$ depends
on the normalization of the gluon polarization. Taking the estimate
$\Delta g_{\rm intrinsic} = +0.5$ at  1~GeV$^2$ \cite{gluon}, we obtain
$\Gamma_c^{\rm int} = -0.0020$ and
$(\Gamma_u + \Gamma_d + \Gamma_s)^{\rm int} = -0.07$ when $Q^2=100~
$GeV$^2$.

It is interesting to compare these estimates of $\Gamma_c$ with the
results of heavy-quark effective theory \cite{manohar} and heavy-quark
operator product expansion \cite{rjc} calculations.  These calculations
express the total heavy-quark contribution to the first moment of $g_1$
in terms of the three-light-flavor singlet axial-charge $g_A^{(0)}$.
Using the most recent value, Eq. (\ref{eq:17}), of $g_A^{(0)}$, Manohar's
effective theory calculation of the heavy-charm-quark axial-charge
becomes
\begin{equation}
g_A^{({\rm charm})} (Q^2 >> m_c^2) = - 0.0055
\pm 0.0018 + O(1/m_c) \ .
\label{eq:19}
\end{equation}
Manohar gives an estimate $\simeq
0.003$ (magnitude) for the $O(1/m_c)$ corrections.

The sum of our extrinsic and intrinsic charm contributions
$(\Gamma_c^{\rm ext} + \Gamma_c^{\rm int} = -0.0044)$ is in good agreement
with Eq. (\ref{eq:19}).  However, with the same gluonic input, photon-gluon
fusion
can account for only about one-third of the difference between the value
of $g_A^{(0)}$ extracted from polarized deep inelastic scattering and
the quark model prediction.  Next-to-leading order QCD fits to the
present world  data for $g_1$ are consistent with a value of $\Delta g$
between zero and +2 at 1 GeV$^2$ \cite{qcdfit}. The value $\Delta g
= 2 $ would increase our estimate of the intrinsic gluon
contribution by a factor of $4$ and would bring theory into agreement
with the empirical determinations of $g_A^{(0)}$.  We look forward
to a more precise measurement of
$\Delta g$ from forthcoming experiments on open charm production.

Finally, it is interesting to note that since the contributions due to
heavy sea quarks come from highly virtual gluons, one expects minimal
nuclear shadowing for their contribution to the first moment of $g_1^N$.

\vspace{2.0cm}

\section*{Acknowledgments}

We thank R. J. Crewther, A. H. Mueller, F. M. Steffens, and A. W. Thomas for helpful
conversations.  This work was supported in part by the United States
Department of Energy under contract number DE--AC03--76SF00515, by
Fondecyt (Chile) under grant 1960536, and by a C\'atedra Presidencial
(Chile).


\newpage

\begin{thebibliography}{99}

\bibitem{efremov}
A.V. Efremov and O.V. Teryaev, JINR Report
E2--88--287 (1988), and in Proceedings of the International Hadron
Symposium, Bechyn\v{e} 1988, eds.\ J. Fischer et al.\
(Czechoslovakian Academy of Science, Prague, 1989) p. 302.

\bibitem{ar}
G. Altarelli and G. G. Ross, Phys. Lett. {\bf B212} (1988) 391.

\bibitem{ccm}
R. D. Carlitz, J. C. Collins, and A. H. Mueller, Phys. Lett. {\bf B214}
(1988) 229.

\bibitem{leader}
M. Anselmino, A. Efremov, and E. Leader,  Phys. Rept. {\bf 261}
(1995) 1; \\
H.-Y. Cheng, Int. J. Mod. Phys. {\bf A11} (1996) 5109.

\bibitem{ej} J. Ellis, and R. L. Jaffe, Phys. Rev. {\bf D9} (1974) 1444;
(E) {\bf D10} (1974) 1669.

\bibitem{adler}
S. L. Adler, Phys. Rev. {\bf 177} (1969) 2426.

\bibitem{bell}
J. S. Bell and R. Jackiw, Nuovo Cimento {\bf 60A} (1969) 47.

\bibitem{bnt}
S. D. Bass, N.N. Nikolaev, and A.W. Thomas,
Adelaide University preprint ADP-133-T80 (1990) unpublished; 
S. D. Bass, Ph.D. thesis (University of Adelaide, 1992).

\bibitem{dhg}
S. D. Drell and A. C. Hearn, Phys. Rev. Lett. {\bf 162} (1966) 1520; \\
S. B. Gerasimov, Yad. Fiz. {\bf 2} (1965) 839.

\bibitem{Alt72} G. Altarelli, N. Cabibbo, and L. Maiani, Phys.
Lett {\bf B40} (1972) 415.

\bibitem{Brod95} S. J. Brodsky and I. Schmidt, Phys. Lett. {\bf
B351} (1995) 344.

\bibitem{bassbs}
S. D. Bass, S. J. Brodsky, and I. Schmidt, Phys. Lett. {\bf 437} (1998) 417.

\bibitem{rizzo} S. J. Brodsky, T. G. Rizzo, and I. Schmidt,
Phys. Rev. {\bf D52} (1995) 4929.

\bibitem{negative}
A. I. Signal and A. W. Thomas,  Phys. Lett. {\bf 191B} (1987) 205;\\
M. Burkardt and  B.  Warr, Phys. Rev. {\bf D45} (1992) 958;\\
S. J. Brodsky  and  B.- Q. Ma,  Phys. Lett.{\bf B381} (1996) 317.

\bibitem{compass} The COMPASS proposal, CERN/SPSLC 96-14.

\bibitem{hermesc}  The HERMES Charm Upgrade Program, HERMES 97-004.

\bibitem{guillet}
J. Ph. Guillet, Z Physik {\bf C39} (1988) 75.

\bibitem{eks}
J. Ellis, M. Karliner and C. T. Sachrajda, Phys. Lett. {\bf B231} (1989) 497.

\bibitem{al}
G. Altarelli and B. Lampe, Z Physik {\bf C47} (1990) 315.

\bibitem{as}
L. Mankiewicz and A. Sch\"afer, Phys. Lett. {\bf 242} (1990) 455.

\bibitem{manohar}
A.V. Manohar, Phys. Lett. {\bf B242} (1990) 94.

\bibitem{grva}
M. Gl\"uck, E. Reya, and W. Vogelsang, Nucl. Phys. {\bf B351} (1991) 579;
\\ W. Vogelsang, Z. Physik {\bf C50} (1991) 275.

\bibitem{bt92}
S. D. Bass and A.W. Thomas, Phys. Lett. {\bf B293} (1992) 457.

\bibitem{steffens}
F. M. Steffens and A. W. Thomas,  Phys. Rev. {\bf D53}, 1191 (1996).

\bibitem{neervan}
W. L. van Neerven, Acta Phys. Polon. {\bf B28} (1997) 2715,
lectures presented at the XXXVIIth Cracow School on Theoretical
Physics, June 1997, Zakopane, Poland, and references therein.

\bibitem{bosted}
P. Bosted, private communication.

\bibitem{grvb}
M. Gl\"uck, E. Reya, and A. Vogt, Z Physik {\bf C67} (1995) 433.

\bibitem{bag}
F. M. Steffens, H. Holtmann, and A.W. Thomas, Phys. Lett. {\bf B358} 
(1995) 139.

\bibitem{parisi}
G. Parisi and R. Petronzio, Nucl. Phys. {\bf B154} (1979) 427.

\bibitem{shore}
G. M. Shore and G. Veneziano, Mod. Phys. Lett. {\bf A8} (1993) 373.

\bibitem{brod97}
S. J. Brodsky and I. Schmidt, Phys. Lett. {\bf B423} (1998) 145.

\bibitem{myhrer}
F. Myhrer and A.W. Thomas, Phys. Rev. {\bf D38} (1988) 1633.

\bibitem{brodin}
G. P. Lepage and S. J. Brodsky, Phys. Rev. {\bf D22} (1980) 2157.

\bibitem{cloudy}
A.W. Schreiber and A.W. Thomas, Phys. Lett. {\bf B215} (1988) 141.

\bibitem{weise}
K. Suzuki and W. Weise, Nucl. Phys. {\bf A634} (1998) 141.

\bibitem{schlumpf}
S. J. Brodsky and F. Schlumpf, Phys. Lett. {\bf 329} (1994) 111.

\bibitem{gluon}
S. J. Brodsky and I. Schmidt, Phys. Lett. {\bf B234} (1990) 144; \\
S. J. Brodsky, M. Burkardt, and I. Schmidt, Nucl. Phys.
{\bf B441} (1995) 197.

\bibitem{extms}
S. J. Brodsky, C.-R. Ji, A. Pang, D. G.
Robertson, Phys. Rev. {\bf D57}, 245 (1998).

\bibitem{alphav}
S. J. Brodsky, G. P. Lepage and P. B. Mackenzie, Phys. Rev. {\bf D28} 
(1983) 228.

\bibitem{expt} EMC Collaboration
(J Ashman et al.) Phys. Lett. {\bf B206} (1988) 364;
Nucl. Phys. {\bf B328} (1989) 1.

\bibitem{smc}
The Spin Muon Collaboration (D. Adams et al.),
Phys. Lett. {\bf B396} (1997) 338; \\
(B. Adeva et al.), Phys. Lett. {\bf B412} (1997) 414;
Phys. Rev. {\bf D58} (1998) 112001

\bibitem{hermes}
The HERMES Collaboration (K. Ackerstaff et al.),
Phys. Lett. {\bf B404} (1997) 383.

\bibitem{slac}
The E-143 Collaboration (K. Abe et al.),
Phys. Rev. Lett. {\bf 74} (1995) 346; Phys. Rev. {\bf D58} (1998) 112003; \\
The E-154 Collaboration (K. Abe et al.), Phys. Rev. Lett. {\bf 79} (1997) 26.

\bibitem{bass98}
S. D. Bass, Mod. Phys. Lett. {\bf A13} (1998) 791.

\bibitem{rjc}
S. D. Bass, R. J. Crewther, F. M. Steffens, and A.W. Thomas, in preparation.



\bibitem{qcdfit}
T. Gehrmann and W.J. Stirling, Phys. Rev. {\bf D53} (1996) 6100; \\
J. Ellis and M.  Karliner, Phys. Lett {\bf B341} (1995) 397;\\
G. Altarelli, R. D. Ball, S. Forte, and G. Ridolfi, Nucl.
Phys. {\bf B496}  (1997) 337; \\
M. Stratmann, hep-ph/9710379; \\
D. de Florian, O. A. Samapayo, and R. Sassot, Phys. Rev {\bf D57} (1998)
5803; \\ L. E. Gordon, M. Goshtasbpour, and G. P. Ramsey,
Phys. Rev. {\bf D58} (1998) 094017; \\
E. Leader, A.V. Sidorov, and D. B. Stamenov, hep-ph/9808248.

\end{thebibliography}
\end{document}